# HABITABILITY OF KNOWN EXOPLANETARY SYSTEMS BASED ON MEASURED STELLAR PROPERTIES

**BARRIE W. JONES, P. NICK SLEEP, AND DAVID R. UNDERWOOD**

**Astronomy Group, The Open University, Milton Keynes, MK7 6AA, UK**

**b.w.jones@open.ac.uk**

# ABSTRACT

At present, because of observational selection effects, we know of no exoplanetary systems with any planetary masses close to that of the Earth. We have therefore used computer models to see whether such planets could be dynamically stable in the presence of the more massive planets known to be present, and in particular whether 'Earths' could remain confined to the classical habitable zone (HZ) for long enough for life to have emerged.

Measured stellar properties have been used to determine for each system the present location of the classical habitable zone (HZ). We have also determined the critical distances from the orbit of each giant planet within which an Earth-mass planet would suffer large orbital changes. We then evaluated the present habitability of each and every exoplanetary system by examining the penetration of these critical distances into the HZ. The critical distances can be obtained by extensive computer modelling of an exoplanetary system. This is far too time consuming to apply to all of the 150 or so systems already known, and to keep up with the latest discoveries. Therefore, in earlier work we studied a few systems in great detail, and developed a speedier means of obtaining the critical distances. We summarize this comparatively quick method here. We can then evaluate comparatively quickly the present habitability of each exoplanetary system by examining the penetration of the critical distance(s) into the HZ.

Of the 152 exoplanetary systems known by 18 April 2006, 60% offer safe havens to Earth-mass planets across >20 % of the HZ width. We regard such systems as being habitable today. We have also estimated whether habitability is possible for 1000 Ma into the past (provided that this period post-dates the heavy bombardment). Of the 143 systems that are susceptible to this second analysis, we find that about 50% offer habitability sustained over 1000 Ma. If giant planets interior to the HZ have got there by migration through the HZ, and if this ruled out the subsequent formation of Earth-mass planets, then 60% and 50% fall to 7% in both cases.

In earlier work (Jones et al. 2005) we used a stellar evolution model (rather than measured stellar properties) to study the 111 systems then known, throughout each star's main-sequence lifetime. For 103 systems we were able to obtain the present-day habitability, and the results are similar in nearly all cases to those in this present work. For two such different approaches, this gives us confidence in each set of results.

# 1. INTRODUCTION

At present, because of observational selection effects, we know of no exoplanetary systems with planetary masses close to that of the Earth – no ‘Earths’. We have therefore used computer models to see whether such planets could be dynamically stable in the presence of the more massive planets (‘giants’) known to be present, and in particular whether ‘Earths’ could remain *confined* to the classical habitable zone (HZ) for long enough for life to have emerged. If so, then it is possible that life is present on any such planets.

The essence of our approach is, first, to establish the location of the HZ around a star. Second, we obtain the critical distances interior and exterior to each of its planets, within which an ‘Earth’ would suffer large orbital changes, ending in ejection or collision. If these distances span the entire HZ then confinement is nowhere possible in the HZ, and the system is classified as uninhabitable. If all, or a significant proportion (>20%) of the HZ, is free from penetration by the critical distances, then confinement is possible over some or all of the HZ and the system is classified as habitable. As a star ages, the HZ migrates outwards and so there is a distinction between instantaneous habitability and habitability for long enough that life could emerge. This we take to be 1000 Ma, excluding the first 700 Ma on the main-sequence to allow for a presumed heavy bombardment (Jones 2004).

To locate the HZ (see Section 2.1 for details) we need the luminosity $L$ of the star and its effective temperature $T_e$. In this present work we have used the measured properties of the stars in the known exoplanetary systems. Clearly this yields their instantaneous habitability *today*. Denoting the present by ‘p’, for $T_e$(p) the required measured properties are spectral type and luminosity class. For $L$(p) we require the stellar distance $d$, the apparent visual magnitude $V$, and the bolometric correction $BC$. $BC$ depends on the spectral type and luminosity class. We can also estimate, from the stellar age, whether life has had time to emerge on any ‘Earths’ that might be present. Stellar ages have been obtained (with considerable uncertainty) from stellar spectra, for example, from chromospheric activity (Donahue 1998).

We have investigated the habitability of the 152 exoplanetary systems known by 18 April 2006. This is of interest in itself, and the use of measured stellar properties for present-day habitability is an obvious choice over the use of a stellar evolution model. In earlier work (Jones et al. 2005) we investigated most of the 111 systems then known to examine habitability over the whole main sequence lifetime, and for this a stellar evolution model was essential. The model used was that of Mazzitelli (1989). The input parameters required differ from those listed above, which only allow us to calculate $T_e$(p) and $L$(p) and not how these change with stellar age. For the model the main input parameters are the stellar mass $M$, its metallicity $Z$, and its helium mass fraction $Y$, all obtained from measurements. The model yields $T_{eM}(t)$ and $L_M(t)$ ('M' for 'Mazzitelli), which will include $T_{eM}$(p) and $L_M$(p) but clearly derived from a very different set of input parameters from that required for calculating $T_e$(p) and $L$(p). We have compared the habitability outcome from both approaches. We would not expect great differences, but this needed to be tested.

# 2. CLASSICAL HABITABLE ZONE AND CRITICAL DISTANCES

### *2.1 Determining the classical habitable zone boundaries*

The classical habitable zone is that range of distances from a star where water at the surface of an Earth-like planet would be in the liquid phase. Habitable zones created, for example, by tidal heating, are not included. We have used boundaries for the HZ derived from the work of Kasting, Whitmire, & Reynolds (1993). The inner boundary is the maximum distance from the star where a runaway greenhouse effect would lead to the evaporation of all surface water, and the outer boundary is the maximum distance at which a cloud-free $CO_2$ atmosphere could maintain a surface temperature of 273K. Alternative criteria have also been applied by Kasting et al. (1993)., which give boundaries to each side of these. Because of simplifications in their climate model, our chosen HZ is conservative in that the real HZ might be wider. At the inner boundary they have neglected enhanced cloud formation, so the runaway greenhouse could occur somewhat closer to the star. At the outer boundary, work by Forget and Pierrehumbert (1997) shows that the formation of $CO_2$ clouds, via a scattering greenhouse effect, might move this boundary outwards. Williams and Kasting (1997) and Mischna et al. (2000) reached a similar conclusion. However, we are encouraged in our choice of boundary criteria by the outcome for the Solar System (Jones 2004 Fig2.4(a)), which matches what we know about Venus, the Earth, and Mars.

To obtain the HZ boundaries we need to use the stellar flux $S_b$ that occurs at each boundary. These have been established by Kasting et al. (1993). This flux depends mainly on $L$, but to some extent on the effective temperature $T_e$ of the star. This is because the lower the value of $T_e$ the greater the infrared fraction in $L$. The greater this fraction, the greater the greenhouse effect for a given stellar flux. Thus, the boundaries are closer to the star than they would have been if $T_e$ had had no effect. We denote the critical flux by $S_b(T_e)$, which in units of the solar constant is given by

$$S_{bi}(T_e) = 4.190\ 10^{-8}\ T_e^2 - 2.139\ 10^{-4}\ T_e + 1.296 \qquad (1a)$$

at the inner boundary, and

$$S_{bo}(T_e) = 6.190\ 10^{-9}\ T_e^2 - 1.319\ 10^{-5}\ T_e + 0.2341 \qquad (1b)$$

at the outer boundary, where $T_e$ is in kelvin. The boundaries are then at distances from the star in AU given by

$$r_i = (L/S_{bi}(T_e))^{1/2} \qquad (2a)$$

$$r_o = (L/S_{bo}(T_e))^{1/2} \qquad (2b)$$

where $L$ is the luminosity of the star in solar units and $S_b(T_e)$ is in units of the solar constant.

In this work we have obtained $L$ and $T_e$ from measured properties of stars. $L$ (in solar units) is obtained from

$$L = 0.787d^2\ 10^{[-0.4(V+BC)]} \qquad (3)$$

where $V$ is the apparent visual magnitude and $BC$ is the bolometric correction (the apparent bolometric magnitude is $(V + BC)$), and the distance $d$ to the star is in parsecs (pc). The spectral type and luminosity class from which the $T_e$ have been obtained are tabulated by Schneider[1], as are $d$ and $V$. The $BC$s are from Cox (2000).

What are the uncertainties in obtaining $L$ and $T_e$ from measured stellar properties in this way? In the calculation of $L$ from eqn(3), the uncertainty is dominated by that in $d$. Many of these distances come from Hipparcos, where the measured parallax has a median standard error of $0.97\times10^{-3}$ arcsec (Perryman et al. 1997). At 100 pc this is ±10%. From eqs(2) and (3) we see that this translates into a ±10% uncertainty in $r$. Values of $T_e$ are perhaps subject to less uncertainty. It is also the case that the $S_b$ are only weakly dependent on $T_e$ (eqn(1)). For example, for our HZ boundary criteria, at around 5700 K, a change of 300 K changes $S_b$ at each boundary by only about 5%. Also, the $r$ values go as the square root of $L$ and $S_b$ (eqn(2)), thus reducing its sensitivity to $L$ and $S_b$ to about half. The uncertainties in $L$ are thus significant but not serious.

### *2.2 Determining a giant planet's critical distances*

With a HZ obtained in this way we need to test to what extent the giant planet(s) known to be present in the system prevent(s) an 'Earth' being confined to the HZ, in the sense that the semimajor axis does not stray outside the HZ. This is determined by the critical distances from the giant(s). On the starward side of a giant planet it is at a distance $n_{int}R_H$ interior to the periastron of the giant planet. On the other side it is at a distance $n_{ext}R_H$ exterior to the apastron. $R_H$ is the Hill radius of a giant planet, defined by

$$R_H = \left(\frac{m_G}{3M_{star}}\right)^{1/3} a_G \qquad (4)$$

where $m_G$ is the mass of the giant planet, $a_G$ is its orbital semimajor axis, and $M_{star}$ is the mass of the star. The critical distances are thus at

- $a_G(1 - e_G) - n_{int}R_H$ interior to the semimajor axis of the giant's orbit
- $a_G(1 + e_G) + n_{ext}R_H$ exterior to the semimajor axis of the giant's orbit

where $e_G$ is the eccentricity of the giant's orbit.

Within these distances confinement is unlikely. If they extend across the whole HZ then nowhere in the HZ offers a safe haven.

Note that even if confined to the HZ, the orbital eccentricity of an 'Earth' will generally increase, and might rise to the point where the planet is carried outside the HZ for a significant fraction of its orbital period. Whether a planet could be habitable in such a case depends on the response time of the atmosphere-ocean system – Williams and Pollard (2002) conclude that a planet like the Earth probably could. For example, if the planet's eccentricity $e \sim 0.2$, the ratio of periastron stellar flux to apastron stellar flux, $[(1 + e)/(1 - e)]^2$, is about the same as the summer/winter flux ratio at mid-latitudes on Earth. They conclude that such an 'Earth' would be habitable as long as its semimajor axis $a$ remained in the HZ, which is our confinement criterion. The habitability limit for $e$ for an 'Earth' is probably between 0.5 and 0.7. For confined orbits we find from orbital integration that $e$ is usually less than about 0.3 and rarely exceeds 0.4.

To obtain the $n_{\rm int}$ and $n_{\rm ext}$ values we have studied in detail seven systems with the MERCURY package of orbital integrators (Chambers 1999). An 'Earth' was launched into a circular orbit at various semimajor axes in the HZ, and its fate followed for a gigayear of simulated time. Our studies of these systems were sufficiently detailed to consume over a thousand hours of CPU time on fast PCs. Full details are in Jones et al. (2005). Key discoveries are as follows.

- $n_{\rm int}$ and $n_{\rm ext}$ are sensitive only to the eccentricity $e_{\rm G}$ of the giant planet's orbit (and not, for example, to $m_{\rm G}/M_{\rm star}$ and $a_{\rm G}$).
- It is an increase in the eccentricity of the orbit of an 'Earth' that leads to its ejection or collision. A giant planet can pump up this eccentricity to large values without more than a few percent change in the semimajor axis of the 'Earth'. (Mean-motion and secular resonances can enhance the pumping.)

The values of $n_{\rm int}(e_{\rm G})$ and $n_{\rm ext}(e_{\rm G})$ are shown in Figure 1, where the data points are connected by cubic fits. These fits to the $n_{\rm int}$ and $n_{\rm ext}$ data are excellent, with correlation coefficients $\rho$ such that $\rho^2 = 0.970$ for $n_{\rm int}$ and 0.998 for $n_{\rm ext}$.

Figure 1 $n_{\rm int}(e_{\rm G})$ and $n_{\rm ext}(e_{\rm G})$ for seven systems (points) studied in detail, and a cubic fit to these points.

The values of $n_{\rm int} = n_{\rm ext} = 3$ at low eccentricity are in accord with analytical values obtained by Gladman (1993) and others for $e_{\rm G}$ close to zero. Analytical solutions are not possible at higher eccentricity. See Jones et al (2005) for a brief discussion about why there is a relationship between $n_{\text{int}}$, $n_{\text{ext}}$, and $e_G$.

[1] Schneider J., http://www.exoplanet.eu/

## 3. HABITABILITY OF THE KNOWN EXOPLANETARY SYSTEMS

Armed with the critical distances from a giant planet we can now see whether this reduces the extent to which the HZ would offer confinement for an 'Earth'. There are six distinct types of configuration, labelled 1-6 in Figure 2, where the lines represent the critical distances from the semimajor axis of the giant. Because the HZ migrates outwards as the star goes through its main-sequence lifetime, these are instantaneous configurations, and will change with stellar age. The confinement outcome relates to the configuration as follows:

| *configuration* | *confinement outcome* | *system habitability today* |
|---|---|---|
| 1, 2 | confinement throughout the HZ | YES |
| 3, 4, 5 | *x*, *y*, *z* %, fraction of HZ width offering confinement | <value>% |
| 6 | confinement nowhere in the HZ | no |

We initially assume that a giant planet interior to the HZ will *not* have ruled out the presence of 'Earths' beyond the giant, even though the giant will probably have got there by traversing the HZ. We then examine the result when the formation of 'Earths' *is* ruled out by such traversal.

Figure 2 Six configurations of the gravitational reach of a giant planet with respect to the instantaneous position of the HZ.

We are interested in two scenarios.

1 The system habitability *today*, which is determined by the confinement outcome today. The possible system habitabilities today are defined above, as 'YES', a percentage, or 'no'.

2 *Sustained* habitability, which requires the star to be at least 1700 Ma old. The first 700 Ma covers a presumed heavy bombardment phase as on Earth and elsewhere in the inner Solar System, followed by at least 1000 Ma for life to emerge, which for the Earth is about the most pessimistic delay (Jones 2004 Section 3.2.1). The possible results are as follows.

- If the system habitability today is 'YES' then the *sustained* habitability is
  - 'YES' if the star's age is ≥ 1700 Ma and it is on the main-sequence, as indicated by luminosity class V
  - 'Too young' if the star's age is < 1700 Ma
  - 'No age' if a plausible age is unavailable (see below)

– ‘Post m-s’ if the star has left the main-sequence, as indicated by a luminosity class IV, III, or (rarely) II.

- If the system habitability today is a % then the *sustained* habitability is
  – ‘YES’ if >20%, if the star’s age is ≥ 1700 Ma, and if it is on the main-sequence as indicated by luminosity class V
  – ‘no’ if ≤ 20%; the case is too marginal, so we take the pessimistic view
  – ‘Too young’, ‘No age’, ‘Post m-s’ as defined above.
- If the system habitability *today* is ‘no’ then the *sustained* habitability outcome is ‘no’.

Stellar ages in the literature are, in most cases, based on spectroscopic data, and we have used these ages. For some of the stars no age has been found in the literature. For most of these cases we have obtained plausible estimates by calculating how the star’s luminosity evolves during the main-sequence, and determine the age at which the present luminosity occurs. The present luminosity is obtained from observed stellar properties, using eqn(3). The evolutionary model is that of Mazzitelli, outlined earlier.

Note that to establish sustained habitability we require an estimate of how much the HZ has moved outwards in the past 1000 Ma. Except in very marginal cases this can be done well enough from the star’s mass for class V and IV stars. Marginal cases are excluded by the ≤ 20% criterion given above.

Table 1 summarises the results of this kind of analysis applied to the 152 exoplanetary systems in Schneider as at 18 April 2006. Note the following.

1 The systems are ordered by increasing period of the planet closest to each star.

2 Under luminosity class, values in brackets () have been inferred by us on the basis of age and mass, and sometimes spectral class also; all are V. In five cases the spectral type sub-divisions (numerical) have been estimated – also shown in brackets.

3 The great majority of systems have been observed only by the radial velocity (Doppler shift) method (Jones 2004). This gives $m_G\sin(i_0)$, where $i_0$ is the inclination of the planetary orbit with respect to the plane of the sky. The minimum actual mass is for $i_0 = 90°$, corresponding to an edge-on presentation of the orbit, and it is these values that are tabulated. However, $i_0$, with a few exceptions (see below), is unknown. Therefore, for the $i_0$-unknown systems the minimum mass has been multiplied by 1.3, this being the reciprocal of the mean value of $\sin(i_0)$, though because $R_H$ varies slowly with $m_G$, as $m_G^{1/3}$ (eqn(1.4)), the 1.3 multiplier increases $R_H$ by only 9% over the minimum mass value, which will make little or no difference to the great majority of habitability results.

4 Minimum masses have been used for the 10 systems shown in *italics*. Of these, 9 have planets that have been observed in transit with sufficient precision that we know $i_0$ to be >81° in all cases, and thus the multiplier is ≥ 0.99. Gliese 876b has been observed astrometrically, and $i_0 \approx$ 90°, again giving a multiplier very close to 1.

5 For systems with more than one giant planet the $nR_H$ for each giant planet is obtained and the confinement outcome is based on the combined gravitational reaches. The configuration in Table 1 is for the giant with the worst (highest numbered) configuration.

6 A few of the stars are components of binary stellar systems. In the closest binaries the stellar separations are about 20 AU, specifically HD 41044A, Gliese 86, and γ Cephei A (Eggenberger et al. 2004). We have not included the gravitational effect of these companion stars, so further study is needed and the results in Table 1 are preliminary, though γ Cephei A is already uninhabitable and will remain so. Studies by David et al. (2003) indicate that binary stars with separations much greater than 20 AU are very unlikely to have their habitability disturbed by the stellar companion, which thus applies to the other presently known binary systems, where the next closest separation is 100AU.

TABLE 1

THE HABITABILITY OF THE KNOWN EXOPLANETARY SYSTEMS,

WITH HZS BASED ON OBSERVED STELLAR PROPERTIES.

| Star name | $M_{star}$/ $M_{Sun}$ | Spec type | BC | *d*/pc | V | HZ inner /AU | HZ outer /AU | Planet | Min mass/ $m_J$ | *a*/AU | *e* | Config | System habitability today | Sustained habitability? |
|---|---|---|---|---|---|---|---|---|---|---|---|---|---|---|

{SEE END OF DOCUMENT}

From Table 1 we have established that 60% of the 152 exoplanetary systems listed offer habitability today – they have ‘today’ outcomes of ‘Yes’ or ‘>20%’. If, however, giant planets interior to the HZ have got there by migrating through the HZ, and if this ruled out the subsequent formation or survival of ‘Earths’ then the proportion falls to only 7%.

To determine *sustained* habitability we need to establish whether the star has an age greater than 1700 Ma. It is fortunate that this is all we need from the age, because estimates of stellar ages are fraught with uncertainty. Even so, for 9 of the systems we have not been able to obtain plausible ages, indicated by ’no age’ in Table 1. Of the remaining 143 systems, the ‘YES’ proportions for sustained habitability are

- 41% if all ‘Post m-s’ stars are excluded (luminosity classes other than V)
- 50% if class IV stars are included (provided that the only reason for exclusion was post m-s)

- 7% if giant planets interior to the HZ have got there by migration through the HZ, and if this ruled out the subsequent formation of 'Earths'

The inclusion of class IV stars is justifiable because most of them will still be evolving fairly sedately, whereas the luminosities of class III and II stars are changing rapidly.

# 4. COMPARISON WITH OTHER WORK

In earlier work of ours (Jones et al. 2005) we used a stellar evolution model, rather than measured stellar properties, to obtain the confinement outcomes for the then known exoplanetary systems with class V and IV stars. Of the 110 then listed, we were able to obtain what we termed 'now' results for 103, the rest having no age estimates that we were able to use. Of the 103, Gliese 777A has since been deleted from the list of confirmed exoplanetary systems. We thus obtained 'now' results for 102 of the 152 systems in Table 1.

We have compared these 'now' results with the 'sustained habitability?' results in the present work for the common set of 102 systems. In essence this is comparing like with like, because in both cases the 'now' results and 'sustained habitability?' results are based on habitability over the past 1000 Ma, and class IV stars are not barred from consideration. On the basis of observed stellar properties, of the 102 systems we get 51% offering sustained habitability, compared to 49% in our earlier work. If the migration of giant planets through the HZ prevented formation or survival of 'Earths', then these proportions are 6% in both cases.

Though this agreement is striking, we examined the 11 individual cases where the 'now' result is different from the 'sustained habitability?' result. For example, if the 'now' result is equivalent to a 'YES', the 'sustained habitability?' result is 'no', and vice versa. (Remembering that 'YES' can correspond to as little as 20% of the HZ offering habitability, these differences are not as stark as they might at first appear.) All of these have been resolved, as follows.

- In 2 cases we have revised the ages of the stars, in one case $\tau$ Bootis from 1500 Ma to 2000 Ma, in the other, Gliese 86 from 3000 Ma to 1000 Ma, thus moving across the 1700 Ma age criterion boundary in each case.
- In 3 cases the system has changed, in 2 cases through new planets being discovered plus attendant revisions to the earlier discovered planets, and in one case by revision of the star's mass. (Revisions in other systems have not changed the results.)

In the remaining 6 cases, the different outcomes are within the discrepancies expected between using a stellar evolution model and using measured stellar properties, each of which is subject to uncertainty.

We can thus conclude that the results of our work regarding sustained habitability in the current exoplanetary system population, are not sensitive to the stellar properties as determined variously by stellar models and observations. This is not unexpected, but it was important be certain. For two such different approaches, this agreement gives us confidence in each set of results.

Note that the exoplanetary systems that have been discovered since the 110 in our earlier work, have not much changed the distribution with planetary semimajor axis, though they have somewhat increased the proportion of giant planets interior to the HZ.

Few others have examined the habitabilites of a large proportion of the known systems. Menou and Tabachnik (2003) studied 85 known exoplanetary systems by scattering particles across the zero-age HZ and examining the number still present in the HZ after 1 Ma. They require that the particle remains in the HZ in the whole of its orbit, and not just that the semimajor axis does so. In spite of the substantial differences between their approach and ours, and their use of zero-age HZs, there is remarkable correspondence between the results at comparable giant planet masses. For example, their cases where a large proportion of the particles was retained, correspond nearly always to 'YES' in 'sustained habitability?' in Table 1, and where a small proportion was retained this corresponds to 'no'.

In a different type of study, Turnbull and Tarter (2003) examined 55 known exoplanetary systems using a technique similar to ours for 'sustained habitability?'. They assumed minimum giant masses and used 3 for $n_{\mathrm{int}}$ and $n_{\mathrm{ext}}$ as the $R_{\mathrm{H}}$ multiplier. The use of $n_{\mathrm{ext}} = 3$ will overestimate the confinement possible when the giant planet is in an eccentric orbit interior to the HZ. Making an allowance for this, and for significant differences in stellar parameters in several systems, there is reasonable agreement between their work and ours.

## 5. CONCLUSIONS AND FUTURE WORK

Of the 152 exoplanetary systems listed by Schneider in 18 April 2006, 143 have ages estimated well enough for us to establish whether they have offered sustained habitability i.e. over the past 1000 Ma, providing that this postdates 700 Ma allowed for a heavy bombardment. Of these 143, we have

- 41% 'YES', and thus have had sustained habitability
- an increase to 50% if class IV stars are included (provided that the only reason for exclusion was post m-s)
- a decrease to 7% if giant planets interior to the HZ have got there by migration through the HZ, and if this ruled out the subsequent formation of 'Earths'.

Comparison with our earlier work reveals that these results are not sensitive to stellar properties as determined variously by stellar models and observations.

The decrease to 7% demonstrates the importance of understanding how readily or rarely at least one 'Earth' can form in the HZ after a giant planet has migrated through it. This urgent question has received some attention. Formation in 47 Ursae Majoris has been examined by Laughlin et al. (2002). They have shown that Earth-mass planets could form within about 0.7 AU of the star, which is interior to the HZ, and possibly a bit further out in the inner HZ. It is the proximity of the inner giant planet to the HZs that hinders formation, by stirring up the orbits of the planetesimals and planetary embryos. Armitage (2003) concluded that post-migration formation of 'Earths' might be unlikely, though he concentrated on the effect of giant migration on planet-forming dust rather than on planetesimals and planetary embryos. On the other hand, Mandell and Sigurdsson (2003) have shown that when the HZ is traversed by a giant planet, a significant fraction of any pre-formed terrestrial planets could survive, eventually returning to circular orbits fairly close to their original positions. An optimistic outcome has also been obtained by Fogg and Nelson (2005), who have shown that post-migration formation of 'Earth' from planetesimals and planetary embryos is fairly likely. Fogg and Nelson's work is the most comprehensive to date, and gives cause for optimism. But this problem needs further study.

Future discoveries are likely to reveal hot Jupiters by the transit method and systems more like ours by Doppler spectroscopy. This is because of observational biases – the closer the planet is to the star the more likely it will be seen in transit, and the longer we observe systems the more likely we are to observe the periodic effects of a giant planet in an orbit at several AU. If 'Earths' can form after migration of a giant through the HZ, then with the giant lying well interior to the HZ, the HZ should have had sustained habitability. Likewise, systems like ours, with the giant planets parked well beyond the HZ, sustained habitability is again likely. So, the proportion of systems with HZs that have had sustained habitability should rise. Countering this trend would be the discovery of additional planets in the known systems. In cases where such planets are close to the HZ, habitability would be reduced or eliminated. Overall, the proportion of Jupiter-mass planets well beyond the HZ, and as yet undiscovered by Doppler spectroscopy, is likely to increase.

An area that needs more investigation is the orbital stability of putative large satellites of giant planets in the habitable zone, where tidal heating is not necessary to produce liquid water. Work so far (Barnes and O'Brien 2002) indicates that Earth-mass satellites could survive in a wide variety of cases.

We are grateful to John Chambers for discussions, and to an anonymous referee for helpful comments.

FIGURE 1

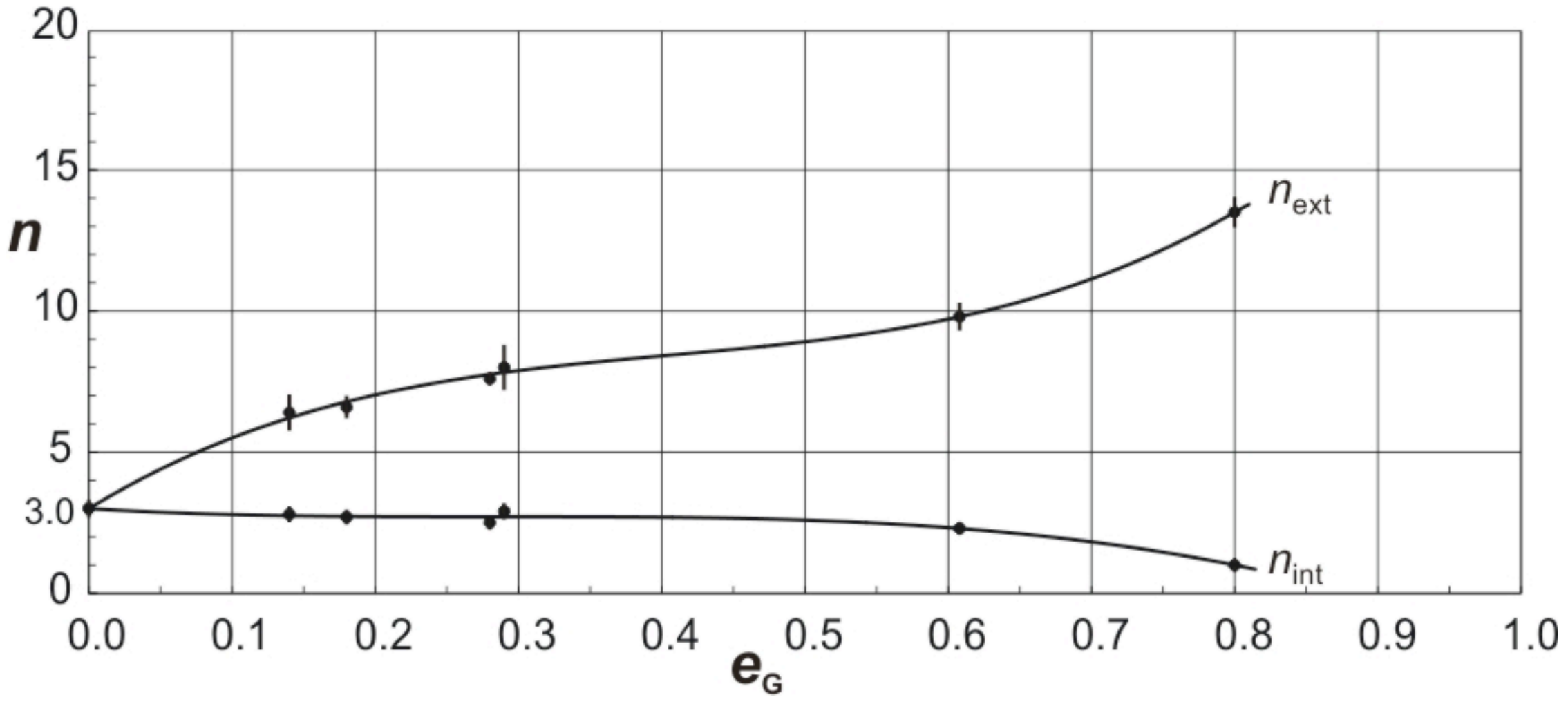

Fig. 1.–$n_{int}(e_G)$ and $n_{ext}(e_G)$ for seven systems (points) studied in detail, and a cubic fit to these points.

FIGURE 2

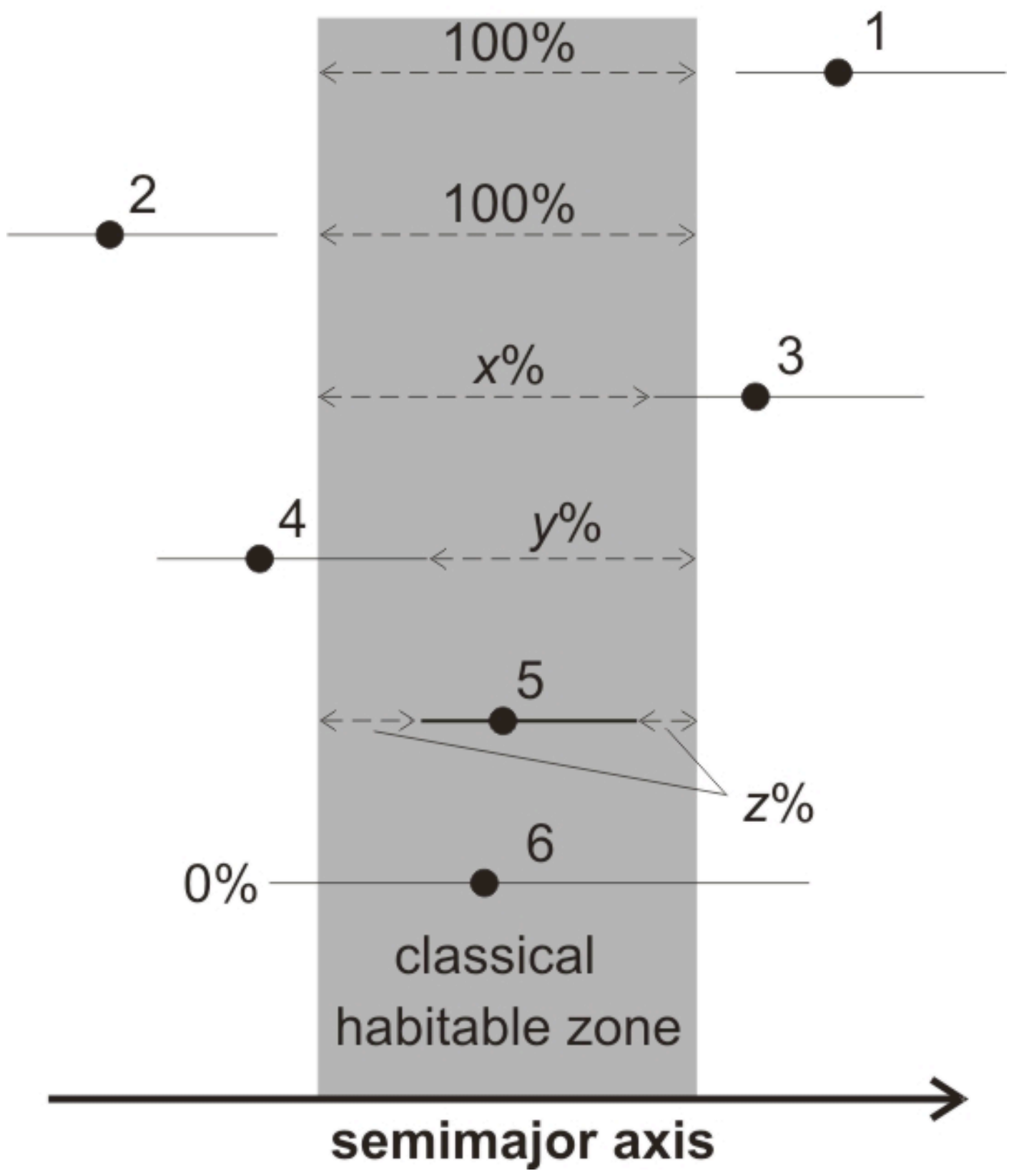

Fig. 2.–Six configurations of the gravitational reach of a giant planet with respect to the instantaneous position of the HZ

**TABLE 1 Habitability outcomes**

| Star name | $M_{star}$/ $M_{Sun}$ | Spec type, class | *BC* | *d*/pc | *V* | HZ inner /AU | HZ outer /AU | Planet | Min mass/ $m_J$ | *a*/AU | *e* | Config | System habita-bility today | Sustained habita-bility? |
|---|---|---|---|---|---|---|---|---|---|---|---|---|---|---|
| *OGLE-TR-56* | 1.04 | G(5)V | -0.18 | 1500 | 16.6 | 0.565 | 1.12423 | b | 1.45 | 0.0225 | 0 | 2 | YES | YES |
| *OGLE-TR-113* | 0.77 | K(5)V | -0.36 | 1500 | 14.42 | 1.832 | 3.600 | b | 1.35 | 0.0228 | 0 | 2 | YES | no age |
| *OGLE-TR-132* | 1.35 | F(5)V | -0.09 | 1500 | 15.72 | 0.713 | 1.452 | b | 1.19 | 0.0306 | 0 | 2 | YES | too young |
| *Gliese 876* | 0.32 | M4V | -2.41 | 4.72 | 10.17 | 0.117 | 0.232 | d | 0.023 | 0.0208 | 0 | 6 | no | no |
| | | | | | | | | c | 0.56 | 0.13 | 0.27 | | | |
| | | | | | | | | b | 1.935 | 0.208 | 0.0249 | | | |
| HD 86081 | 1.21 | F8V | -0.17 | 91 | 8.73 | 1.251 | 2.497 | b | 1.5 | 0.039 | 0.008 | 2 | YES | no age |
| *HD 189733* | *0.83* | *K1-2(V)* | *-0.41* | *19.3* | *7.67* | *0.552* | *1.085* | *b* | *1.15* | *0.0313* | *0* | *2* | *YES* | *YES* |
| HD 212301 | 1.03 | F8V | -0.17 | 52.7 | 7.77 | 1.127 | 2.250 | b | 0.45 | 0.036 | 0 | 2 | YES | too young |
| HD 73256 | 1.05 | G8(V) | -0.30 | 36.5 | 8.08 | 0.787 | 1.549 | b | 1.87 | 0.037 | 0.03 | 2 | YES | too young |
| GJ 436 | 0.41 | M3(V) | -1.98 | 10.2 | 10.68 | 0.162 | 0.321 | b | 0.067 | 0.0278 | 0.12 | 2 | YES | YES |
| 55 Cancri | 1.03 | G8V | -0.30 | 13.4 | 5.95 | 0.770 | 1.517 | e | 0.045 | 0.038 | 0.174 | 2 | YES | YES |
| | | | | | | | | b | 0.784 | 0.115 | 0.0197 | | | |
| | | | | | | | | c | 0.217 | 0.24 | 0.44 | | | |
| | | | | | | | | d | 3.92 | 5.257 | 0.327 | | | |
| HD 63454 | 0.8 | K4V | -0.60 | 35.8 | 9.37 | 0.518 | 1.016 | b | 0.38 | 0.036 | 0 | 2 | YES | YES |
| *HD 149026* | 1.3 | G0IV | -0.18 | 78.9 | 8.15 | 1.456 | 2.896 | b | 0.36 | 0.042 | 0 | 2 | YES | post m-s |
| HD 83443 | 0.79 | K0V | -0.36 | 43.54 | 8.23 | 0.920 | 1.808 | b | 0.41 | 0.04 | 0.08 | 2 | YES | YES |
| HD 46375 | 1.00 | K1IV | -0.48 | 33.4 | 7.94 | 0.858 | 1.685 | b | 0.249 | 0.041 | 0.04 | 2 | YES | post m-s |
| *TrES-1* | 0.87 | K0V | -0.36 | 157 | 11.79 | 0.644 | 1.265 | - | 0.61 | 0.0393 | 0.135 | 2 | YES | YES |
| HD 179949 | 1.24 | F8V | -0.17 | 27 | 6.25 | 1.163 | 2.322 | b | 0.98 | 0.04 | 0.05 | 2 | YES | YES |
| HD 187123 | 1.06 | G5(V) | -0.24 | 50 | 7.86 | 1.129 | 2.231 | b | 0.52 | 0.042 | 0.03 | 2 | YES | YES |
| *OGLE-TR-10* | 1.22 | G(9)(V) | -0.24 | 1500 | 14.93 | 1.306 | 2.580 | b | 0.54 | 0.04162 | 0 | 2 | YES | YES |
| τ Bootis | 1.3 | F7V | -0.16 | 15 | 4.5 | 1.426 | 2.851 | b | 3.9 | 0.046 | 0.01 | 2 | YES | YES |
| HD 188753A | 1.06 | K0(V) | -0.36 | 44.82 | 7.43 | 1.369 | 2.690 | b | 1.14 | 0.0446 | 0 | 2 | YES | YES |
| HD 330075 | 0.95 | G5(V) | -0.24 | 50.2 | 9.36 | 0.568 | 1.123 | b | 0.76 | 0.043 | 0 | 2 | YES | YES |
| HD 88133 | 1.2 | G5IV | -0.29 | 74.5 | 8.01 | 1.647 | 3.244 | b | 0.22 | 0.047 | 0 | 2 | YES | post m-s |
| HD 2638 | 0.93 | G5(V) | -0.24 | 53.71 | 9.44 | 0.586 | 1.158 | b | 0.48 | 0.044 | 0 | 2 | YES | YES |
| BD-10 3166 | 1.1 | G4V | -0.22 | <200 | 10.08 | <1.601 | <3.166 | b | 0.48 | 0.046 | 0 | 2 | YES | no age |
| HD 75289 | 1.05 | G0V | -0.18 | 28.94 | 6.35 | 1.224 | 2.434 | b | 0.42 | 0.046 | 0.054 | 2 | YES | YES |
| *HD 209458* | 1.05 | G0V | -0.18 | 47 | 7.65 | 1.092 | 2.172 | b | 0.69 | 0.045 | 0 | 2 | YES | YES |
| HD 76700 | 1.00 | G6V | -0.25 | 59.7 | 8.13 | 1.210 | 2.388 | b | 0.197 | 0.049 | 0 | 2 | YES | YES |
| *OGLE-TR-111* | 0.82 | G(9)(V) | -0.24 | 1500 | 15.55 | 1.004 | 1.984 | b | 0.53 | 0.047 | 0 | 2 | YES | no age |
| HD 149143 | 1.21 | G0IV | -0.18 | 63 | 7.9 | 1.305 | 2.595 | b | 1.33 | 0.053 | 0.016 | 2 | YES | too young |
| HD 102195 | 0.928 | K0V | -0.36 | 28.98 | 8.05 | 0.688 | 1.35 | b | 0.48 | 0.049 | 0.06 | 2 | YES | YES |
| 51 Pegasi | 1.0 | G2IV | -0.20 | 14.7 | 5.49 | 0.948 | 1.880 | b | 0.468 | 0.052 | 0 | 2 | YES | post m-s |
| υ Andromedae | 1.3 | F8V | -0.17 | 13.47 | 4.09 | 1.569 | 3.132 | b | 0.69 | 0.059 | 0.012 | 6 | no | no |
| | | | | | | | | c | 1.89 | 0.829 | 0.28 | | | |
| | | | | | | | | d | 3.75 | 2.53 | 0.27 | | | |
| HD 49674 | 1.00 | G5V | -0.24 | 40.7 | 8.10 | 0.823 | 1.626 | b | 0.12 | 0.0568 | 0 | 2 | YES | YES |
| HD 109749 | 1.2 | G3IV | -0.21 | 59 | 8.1 | 1.159 | 2.296 | b | 0.28 | 0.0635 | 0.01 | 2 | YES | post m-s |
| Gliese 581 | 0.31 | M3(V) | -2.11 | 6.26 | 10.55 | 0.113 | 0.224 | b | 0.056 | 0.041 | 0 | 2 | YES | YES |
| HD 118203 | 1.23 | K0(V) | -0.36 | 88.6 | 8.05 | 2.034 | 3.997 | b | 2.13 | 0.07 | 0.309 | 2 | YES | post m-s |
| HD 68988 | 1.2 | G0(V) | -0.18 | 58 | 8.21 | 1.041 | 2.071 | b | 1.9 | 0.071 | 0.14 | 2 | YES | YES |
| HD 168746 | 0.92 | G5(V) | -0.24 | 43.12 | 7.95 | 0.934 | 1.846 | b | 0.23 | 0.065 | 0.081 | 2 | YES | YES |
| HD 217107 | 0.98 | G8IV | -0.36 | 37 | 6.16 | 2.024 | 3.977 | b | 1.37 | 0.074 | 0.13 | 6 | no | no |
| | | | | | | | | c | 2.1 | 4.3 | 0.55 | | | |
| HD 162020 | 0.7 | K2V | -0.46 | 31.26 | 9.18 | 0.455 | 0.892 | b | 13.75 | 0.072 | 0.277 | 2 | YES | YES |
| HD 160691 | 1.08 | G3IV/V | -0.21 | 15.3 | 5.15 | 1.169 | 2.316 | d | 0.044 | 0.09 | 0 | 6 | no | no |
| | | | | | | | | b | 1.67 | 1.5 | 0.31 | | | |
| | | | | | | | | c | 3.1 | 4.17 | 0.57 | | | |
| HD 130322 | 0.79 | K0V | -0.36 | 30 | 8.05 | 0.689 | 1.353 | b | 1.08 | 0.088 | 0.048 | 2 | YES | too young |
| HD 108147 | 1.27 | F9V | -0.09 | 38.57 | 6.99 | 1.021 | 2.080 | b | 0.4 | 0.104 | 0.498 | 2 | YES | YES |
| HD 38529 | 1.39 | G4IV | -0.28 | 42.43 | 5.94 | 2.413 | 4.757 | b | 0.78 | 0.129 | 0.29 | 6 | no | no |
| | | | | | | | | c | 12.7 | 3.68 | 0.36 | | | |
| HD 4308 | 0.83 | G5V | -0.24 | 21.9 | 6.54 | 0.908 | 1.794 | b | 0.047 | 0.114 | 0 | 2 | YES | YES |
| Gliese 86 | 0.79 | K1V | -0.41 | 11 | 6.17 | 0.618 | 1.214 | b | 4.01 | 0.11 | 0.046 | 2 | YES | too young |
| HD 99492 | 0.78 | K2V | -0.46 | 18 | 7.57 | 0.549 | 1.078 | b | 0.122 | 0.119 | 0.05 | 2 | YES | YES |
| HD 190360 | 0.96 | G6IV | -0.30 | 15.89 | 5.71 | 1.027 | 2.022 | c | 0.057 | 0.128 | 0.01 | 3 | 64% | YES |
| | | | | | | | | b | 1.502 | 3.92 | 0.36 | | | |
| HD 27894 | 0.75 | K2V | -0.46 | 42.37 | 9.36 | 0.567 | 1.113 | b | 0.62 | 0.122 | 0.049 | 2 | YES | YES |
| HD 33283 | 1.24 | G3V | -0.21 | 86 | 8.05 | 1.729 | 3.424 | b | 0.33 | 0.122 | 0.049 | 2 | YES | no age |
| HD 195019 | 1.02 | G3IV/V | -0.36 | 20 | 6.91 | 0.773 | 1.520 | b | 3.43 | 0.14 | 0.05 | 2 | YES | YES |
| HD 102117 | 0.95 | G6V | -0.25 | 42 | 7.47 | 1.154 | 2.277 | b | 0.14 | 0.149 | 0 | 2 | YES | YES |
| HD 6434 | 1.00 | G3IV | -0.21 | 40.32 | 7.72 | 0.944 | 1.869 | b | 0.48 | 0.15 | 0.3 | 2 | YES | post m-s |
| HD 192263 | 0.79 | K2V | -0.46 | 19.9 | 7.79 | 0.549 | 1.077 | b | 0.72 | 0.15 | 0 | 2 | YES | too young |
| HD 224693 | 1.33 | G2IV | -0.30 | 94 | 8.23 | 1.872 | 3.711 | b | 0.71 | 0.233 | 0.05 | 2 | YES | Post m-s |
| HD 11964 | 1.125 | G5(V) | -0.29 | 33.98 | 6.42 | 1.562 | 3.077 | b | 0.11 | 0.229 | 0.15 | 2 | YES | YES |
| ρ Cor Borealis | 0.95 | G0V | -0.18 | 16.7 | 5.4 | 1.094 | 2.175 | b | 1.04 | 0.22 | 0.04 | 2 | YES | YES |
| HD 74156 | 1.05 | G0(V) | -0.18 | 64.56 | 7.62 | 1.521 | 3.025 | b | 1.86 | 0.294 | 0.636 | 6 | no | no |
| | | | | | | | | c | 6.17 | 3.4 | 0.583 | | | |
| HD 117618 | 1.05 | G2V | -0.20 | 38 | 7.18 | 1.125 | 2.232 | b | 0.19 | 0.28 | 0.39 | 2 | YES | YES |
| HD 37605 | 0.8 | K0V | -0.36 | 42.9 | 8.69 | 0.733 | 1.441 | b | 2.3 | 0.25 | 0.677 | 2 | YES | no age |
| HD 168443 | 1.01 | G5(V) | -0.24 | 33 | 6.92 | 1.149 | 2.270 | b | 7.2 | 0.29 | 0.529 | 6 | no | no |
| | | | | | | | | c | 17.1 | 2.87 | 0.228 | | | |
| HD 3651 | 0.79 | K0V | -0.36 | 11 | 5.8 | 0.712 | 1.398 | b | 0.2 | 0.284 | 0.63 | 2 | YES | no age |
| HD 121504 | 1.00 | G2V | -0.20 | 44.37 | 7.54 | 1.113 | 2.208 | b | 0.89 | 0.32 | 0.13 | 2 | YES | YES |
| HD 101930 | 0.74 | K1V | -0.41 | 30.49 | 8.21 | 0.670 | 1.315 | b | 0.3 | 0.302 | 0.11 | 2 | YES | no age |
| HD 178911 B | 0.87 | G5(V) | -0.24 | 46.73 | 7.98 | 0.999 | 1.973 | b | 6.292 | 0.32 | 0.1243 | 2 | YES | YES |
| HD 16141 | 1.00 | G5IV | -0.29 | 35.9 | 6.78 | 1.398 | 2.755 | b | 0.23 | 0.35 | 0.21 | 2 | YES | post m-s |
| HD 114762 | 0.82 | F9V | -0.17 | 28 | 7.30 | 0.744 | 1.484 | b | 11.02 | 0.3 | 0.34 | 4 | 89% | YES |
| HD 80606 | 0.9 | G5(V) | -0.24 | 58.38 | 8.93 | 0.806 | 1.591 | b | 3.41 | 0.439 | 0.927 | 6 | no | no |
| 70 Virginis | 1.1 | G4V | -0.22 | 22 | 5.00 | 1.827 | 3.614 | b | 7.44 | 0.48 | 0.4 | 2 | YES | YES |
| HD 216770 | 0.9 | K1V | -0.41 | 38 | 8.10 | 0.878 | 1.724 | b | 0.65 | 0.46 | 0.37 | 2 | YES | YES |
| HD 52265 | 1.13 | G0V | -0.18 | 28 | 6.30 | 1.211 | 2.409 | b | 1.13 | 0.49 | 0.29 | 2 | YES | YES |
| HD 208487 | 0.95 | G2V | -0.20 | 45 | 7.48 | 1.161 | 2.302 | b | 0.45 | 0.49 | 0.32 | 2 | YES | YES |
| HD 34445 | 1.11 | G0(V) | -0.18 | 48 | 7.32 | 1.298 | 2.582 | b | 0.58 | 0.51 | 0.4 | 2 | YES | YES |
| GJ 3021 | 0.9 | G6V | -0.24 | 17.62 | 6.59 | 0.714 | 1.411 | b | 3.32 | 0.49 | 0.505 | 4 | 24% | too young |
| HD 93083 | 0.7 | K3V | -0.53 | 28.9 | 8.3 | 0.655 | 1.285 | b | 0.37 | 0.477 | 0.14 | 4 | 90% | No age |
| HD 37124 | 0.91 | G4V | -0.22 | 33 | 7.68 | 0.798 | 1.578 | b | 0.61 | 0.53 | 0.055 | 3 | 44% | YES |
| | | | | | | | | c | 0.6 | 1.64 | 0.14 | | | |
| | | | | | | | | d | 0.66 | 3.19 | 0.2 | | | |
| HD 219449 | 1.7 | K0III | -0.50 | 45 | 4.21 | 6.679 | 13.098 | b | 2.9 | 0.3 | 0 | 2 | YES | post m-s |
| HD 73526 | 1.02 | G6V | -0.25 | 99 | 9.00 | 1.344 | 2.653 | b | 2.9 | 0.66 | 0.19 | 4 | 63% | YES |
| | | | | | | | | c | 2.5 | 1.05 | 0.14 | | | |
| HD 104985 | 1.5 | G9III | -0.46 | 102 | 5.79 | 7.105 | 13.938 | b | 6.3 | 0.78 | 0.03 | 2 | YES | post m-s |
| HD 82943 | 1.05 | G0(V) | -0.18 | 27.46 | 6.54 | 1.064 | 2.116 | c | 0.88 | 0.73 | 0.54 | 6 | no | no |
| | | | | | | | | b | 1.63 | 1.16 | 0.41 | | | |
| HD 169830 | 1.4 | F8V | -0.17 | 36.32 | 5.90 | 1.839 | 3.669 | b | 2.88 | 0.81 | 0.31 | 6 | no | no |
| | | | | | | | | c | 4.04 | 3.6 | 0.33 | | | |
| HD 8574 | 1.15 | F8(V) | -0.17 | 44.15 | 7.12 | 1.274 | 2.543 | b | 2.23 | 0.76 | 0.4 | 4 | 72% | YES |
| HD 202206 | 1.15 | G6V | -0.25 | 46.34 | 8.08 | 0.961 | 1.897 | b | 17.4 | 0.83 | 0.435 | 6 | no | no |
| | | | | | | | | c | 2.44 | 2.55 | 0.267 | | | |
| HD 89744 | 1.4 | F7V | -0.16 | 40 | 5.74 | 2.148 | 4.295 | b | 7.99 | 0.89 | 0.67 | 4 | 70% | YES |
| HD 134987 | 1.05 | G5V | -0.24 | 25 | 6.45 | 1.081 | 2.135 | b | 1.58 | 0.78 | 0.24 | 4 | 67% | YES |
| HD 12661 | 1.07 | G6V | -0.25 | 37.16 | 7.44 | 1.035 | 2.042 | b | 2.3 | 0.83 | 0.35 | 6 | no | no |
| | | | | | | | | c | 1.57 | 2.56 | 0.2 | | | |
| HD 150706 | 1.06 | G0(V) | -0.18 | 27.2 | 7.03 | 0.841 | 1.672 | b | 1 | 0.82 | 0.38 | 4 | 7% | no |
| HD 40979 | 1.08 | F8V | -0.17 | 33.3 | 6.74 | 1.145 | 2.285 | b | 3.32 | 0.811 | 0.23 | 4 | 50% | YES |
| HD 59686 | 1.7 | K2III | -0.60 | 92 | 5.45 | 8.197 | 16.076 | b | 5.25 | 0.911 | 0 | 2 | YES | post m-s |
| HR 810 | 1.2 | G0V | -0.18 | 15.5 | 5.4 | 1.015 | 2.019 | b | 1.94 | 0.91 | 0.24 | 4 | 33% | too young |
| HD 142 | 1.1 | G1IV | -0.19 | 20.6 | 5.70 | 1.190 | 2.364 | b | 1 | 0.98 | 0.38 | 4 | 38% | post m-s |
| HD 92788 | 1.06 | G5(V) | -0.24 | 32.82 | 7.31 | 0.955 | 1.886 | b | 3.86 | 0.97 | 0.27 | 6 | no | no |
| HD 28185 | 0.99 | G5(V) | -0.24 | 39.4 | 7.81 | 0.911 | 1.799 | b | 5.7 | 1.03 | 0.07 | 4 | 1% | no |
| HD 196885 | 1.27 | F8IV | -0.17 | 33 | 6.39 | 1.333 | 2.660 | b | 1.84 | 1.12 | 0.3 | 4 | 34% | post m-s |
| HD 142415 | 1.03 | G1V | -0.19 | 34.2 | 7.34 | 0.929 | 1.844 | b | 1.62 | 1.05 | 0.5 | 6 | no | no |
| HD 33564 | 1.25 | F6V | -0.15 | 20.98 | 5.08 | 1.508 | 3.012 | b | 9.1 | 1.1 | 0.34 | 4 | 18% | no |
| HD 177830 | 1.17 | K0(V) | -0.35 | 59 | 7.17 | 2.031 | 3.991 | b | 1.28 | 1 | 0.43 | 4 | 95% | YES |
| HD 108874 | 1.00 | G5(V) | -0.24 | 68.5 | 8.76 | 1.022 | 2.019 | b | 1.36 | 1.051 | 0.07 | 6 | no | no |
| | | | | | | | | c | 1.018 | 2.68 | 0.25 | | | |
| HD 154857 | 1.17 | G5V | -0.24 | 68.5 | 7.25 | 2.049 | 4.048 | b | 1.8 | 1.11 | 0.51 | 4 | 76% | YES |
| HD 4203 | 1.06 | G5(V) | -0.24 | 77.5 | 8.68 | 1.200 | 2.370 | b | 1.65 | 1.09 | 0.46 | 6 | no | no |
| HD 27442 | 1.2 | K2IV | -0.53 | 18.1 | 4.44 | 2.451 | 4.809 | b | 1.28 | 1.18 | 0.07 | 2 | YES | post m-s |
| HD 210277 | 0.99 | G0(V) | -0.18 | 22 | 6.63 | 0.818 | 1.626 | b | 1.24 | 1.097 | 0.45 | 6 | no | no |
| HD 128311 | 0.8 | K0(V) | -0.36 | 16.6 | 7.51 | 0.489 | 0.960 | b | 2.18 | 1.1 | 0.25 | 3 | 11% | no |
| | | | | | | | | c | 3.21 | 1.76 | 0.17 | | | |
| HD 19994 | 1.35 | F8V | -0.17 | 22.38 | 5.07 | 1.660 | 3.313 | b | 2 | 1.3 | 0.2 | 4 | 59% | YES |
| HD 188015 | 1.09 | G5IV | -0.29 | 52.6 | 8.22 | 1.006 | 1.988 | b | 1.26 | 1.19 | 0.15 | 4 | 6% | no |
| HD 13189 | 4.5 | K2II | -0.60 | 185 | 7.57 | 6.210 | 12.178 | b | 14 | 1.85 | 0.28 | 2 | YES | post m-s |
| HD 20367 | 1.04 | G0(V) | -0.18 | 27 | 6.41 | 1.110 | 2.209 | b | 1.07 | 1.25 | 0.23 | 4 | 1% | no |
| HD 114783 | 0.92 | K0(V) | -0.36 | 22 | 7.57 | 0.630 | 1.238 | b | 0.99 | 1.2 | 0.1 | 3 | 29% | YES |
| HD 147513 | 0.92 | G3V | -0.21 | 12.9 | 5.37 | 0.891 | 1.765 | b | 1 | 1.26 | 0.52 | 6 | no | no |
| HIP 75458 | 1.05 | K2III | -0.60 | 31.5 | 3.31 | 7.520 | 14.747 | b | 8.64 | 1.34 | 0.71 | 2 | YES | post m-s |
| HD 222582 | 1.00 | G5(V) | -0.24 | 42 | 7.70 | 1.021 | 2.017 | b | 5.11 | 1.35 | 0.76 | 6 | no | no |
| HD 20782 | 1.00 | G2V | -0.20 | 36.02 | 7.38 | 0.973 | 1.929 | b | 1.8 | 1.36 | 0.92 | 6 | no | no |
| HD 65216 | 0.92 | G5V | -0.24 | 34.3 | 7.98 | 0.733 | 1.448 | b | 1.21 | 1.37 | 0.41 | 6 | no | no |
| HD 183263 | 1.17 | G2IV | -0.20 | 53 | 7.86 | 1.147 | 2.276 | b | 3.69 | 1.52 | 0.38 | 6 | no | no |
| HD 141937 | 1.00 | G2V | -0.20 | 33.46 | 7.25 | 0.959 | 1.903 | b | 9.7 | 1.52 | 0.41 | 6 | no | no |
| HD 41004 A | 0.7 | K1V | -0.41 | 42.5 | 8.65 | 0.762 | 1.497 | b | 2.3 | 1.31 | 0.39 | 6 | no | no |
| HD 11977 | 1.91 | G8.5III | -0.44 | 66.5 | 4.7 | 7.548 | 14.810 | b | 6.54 | 1.93 | 0.4 | 2 | YES | post m-s |
| HD 47536 | 1.1 | K0II | -0.50 | 123 | 5.26 | 11.256 | 22.074 | b | 4.96 | 1.61 | 0.2 | 2 | YES | post m-s |
| HD 23079 | 1.1 | F9V | -0.17 | 34.8 | 7.1 | 1.028 | 2.048 | b | 2.61 | 1.65 | 0.1 | 6 | no | no |
| 16 Cygni B | 1.01 | G2.5V | -0.20 | 21.4 | 6.20 | 1.002 | 1.985 | b | 1.69 | 1.67 | 0.67 | 6 | no | no |
| HD 4208 | 0.93 | G5V | -0.24 | 33.9 | 7.79 | 0.791 | 1.562 | b | 0.8 | 1.67 | 0.05 | 3 | 57% | YES |
| HD 114386 | 0.75 | K3V | -0.53 | 28 | 8.73 | 0.521 | 1.021 | b | 0.99 | 1.62 | 0.28 | 3 | 63% | YES |
| HD 45350 | 1.02 | G5IV | -0.29 | 49 | 7.88 | 1.150 | 2.265 | b | 0.98 | 1.77 | 0.78 | 6 | no | no |
| γ Cephei A | 1.59 | K0III | -0.60 | 11.8 | 3.22 | 2.936 | 5.758 | b | 1.59 | 2.03 | 0.2 | 4 | 80% | post m-s |
| HD 213240 | 1.22 | G4IV | -0.28 | 40.75 | 6.80 | 1.477 | 2.922 | b | 4.5 | 2.03 | 0.45 | 6 | no | no |
| HD 187085 | 1.22 | G0V | -0.18 | 44.98 | 7.22 | 1.274 | 2.534 | b | 0.75 | 2.05 | 0.47 | 6 | no | no |
| HD 81040 | 0.96 | G2-3(V) | -0.20 | 32.56 | 7.72 | 0.757 | 1.500 | b | 6.86 | 1.94 | 0.526 | 6 | no | no |
| HD 10647 | 1.07 | F8V | -0.17 | 17.3 | 5.52 | 1.043 | 2.082 | b | 0.91 | 2.1 | 0.18 | 3 | 30% | YES |
| HD 10697 | 1.1 | G5IV | -0.29 | 30 | 6.29 | 1.464 | 2.885 | b | 6.12 | 2.13 | 0.11 | 6 | no | no |
| 47 U Majoris | 1.03 | G0V | -0.18 | 13.3 | 5.10 | 1.000 | 1.989 | b | 2.54 | 2.09 | 0.061 | 3 | 34% | YES |
| | | | | | | | | c | 0.76 | 3.73 | 0.1 | | | |
| HD 190228 | 1.3 | G5IV | -0.29 | 66.11 | 7.3 | 2.027 | 3.992 | b | 4.99 | 2.31 | 0.43 | 6 | no | no |
| HD 114729 | 0.93 | G3V | -0.21 | 35 | 6.69 | 1.316 | 2.607 | b | 0.82 | 2.08 | 0.31 | 6 | no | no |
| HD 111232 | 0.78 | G8V | -0.30 | 29 | 7.61 | 0.776 | 1.528 | b | 6.8 | 1.97 | 0.2 | 3 | 6% | no |
| HD 2039 | 0.98 | G2V | -0.20 | 89.8 | 9.01 | 1.145 | 2.270 | b | 4.85 | 2.19 | 0.68 | 6 | no | no |
| HD 136118 | 1.24 | F9V | -0.17 | 52.3 | 6.94 | 1.663 | 3.313 | b | 11.9 | 2.3 | 0.37 | 6 | no | no |
| HD 50554 | 1.1 | F8(V) | -0.17 | 31.03 | 6.86 | 1.010 | 2.015 | b | 4.9 | 2.38 | 0.42 | 6 | no | no |
| HD 196050 | 1.1 | G3V | -0.21 | 46.9 | 7.50 | 1.215 | 2.406 | b | 3 | 2.5 | 0.28 | 6 | no | no |
| HD 216437 | 1.07 | G4V | -0.22 | 26.5 | 6.06 | 1.351 | 2.672 | b | 2.1 | 2.7 | 0.34 | 6 | no | no |
| HD 216435 | 1.25 | G0V | -0.18 | 33.3 | 6.03 | 1.631 | 3.245 | b | 1.49 | 2.7 | 0.34 | 6 | no | no |
| HD 106252 | 1.05 | G0(V) | -0.18 | 37.44 | 7.36 | 0.994 | 1.977 | b | 6.81 | 2.61 | 0.54 | 6 | no | no |
| HD 23596 | 1.3 | F8(V) | -0.17 | 52 | 7.24 | 1.420 | 2.834 | b | 7.19 | 2.72 | 0.314 | 6 | no | no |
| 14 Herculis | 1.00 | K0V | -0.36 | 18.1 | 6.67 | 0.784 | 1.541 | b | 4.74 | 2.8 | 0.338 | 3 | 26% | YES |
| HD 142022 A | 0.99 | K0V | -0.36 | 35.87 | 7.70 | 0.967 | 1.901 | b | 4.4 | 2.8 | 0.57 | 6 | no | no |
| HD 39091 | 1.1 | G1IV | -0.19 | 20.55 | 5.67 | 1.204 | 2.391 | b | 10.35 | 3.29 | 0.62 | 6 | no | no |
| HD 70642 | 1.00 | G5V | -0.24 | 29 | 7.18 | 0.896 | 1.770 | b | 2 | 3.3 | 0.1 | 1 | YES | YES |
| HD 33636 | 0.99 | G0V | -0.18 | 28.7 | 7.06 | 0.875 | 1.740 | b | 9.28 | 3.56 | 0.53 | 6 | no | no |
| ε Eridani | 0.8 | K2V | -0.46 | 3.2 | 3.73 | 0.572 | 1.123 | b | 0.86 | 3.3 | 0.608 | 3 | 17% | no |
| HD 50499 | 1.27 | G1V | -0.19 | 47.26 | 7.22 | 1.356 | 2.693 | b | 1.71 | 3.86 | 0.23 | 3 | 61% | YES |
| HD 117207 | 1.04 | G8V | -0.30 | 33 | 7.26 | 1.038 | 2.043 | b | 2.06 | 3.78 | 0.16 | 1 | YES | YES |
| HD 30177 | 0.95 | G8V | -0.30 | 55 | 8.41 | 1.018 | 2.005 | b | 9.17 | 3.86 | 0.3 | 3 | 16% | no |
| HD 89307 | 1.27 | G0V | -0.18 | 33 | 7.06 | 1.006 | 2.001 | b | 2.73 | 4.15 | 0.27 | 1 | YES | too young |
| HD 72659 | 0.95 | G0V | -0.18 | 51.4 | 7.48 | 1.292 | 2.569 | b | 2.96 | 4.16 | 0.2 | 3 | 71% | YES |

(1) The observed magnitudes of the OGLEs are I not V, except for OGLE-TR-56, where it is V. The appropriate magnitude has been used to calculate *L*.

(2) A multiplier of 1.3 has been used to obtain the final three columns, except for cases (*italicized*) where *i* is known, always >81°, so the minimum mass is used.